\documentclass{appolb}

\usepackage{graphicx}
\usepackage{subfigure}
\usepackage{cite}

\newcommand{\MeV}{\textrm{MeV}}

\begin{document}

\title{Excited Hadrons, Heavy Quarks and QCD
  thermodynamics\thanks{Presented by ERA at {\em Excited QCD 2013},
    3-9 February 2013 Bjelasnica Mountain, Sarajevo,
    Bosnia-Herzegovina.}  \thanks{Supported by Spanish DGI grant
    FIS2011-24149, Junta de Andaluc{\'\i}a grant
    FQM225, FPA2011-25948 and the JdC Program of the
    Spanish MICINN.}} \author{ E. Ruiz Arriola, L.L. Salcedo
  \address{Departamento de F\'{\i}sica At\'omica, Molecular y Nuclear
    \\ and Instituto Carlos I de F{\'\i}sica Te\'orica y
    Computacional, \\ Universidad de Granada, E-18071 Granada, Spain}
  \and E. Megias \address{Grup de F\'\i sica Te\`orica and IFAE,
    Departament de F\'\i sica, \\ Universitat Aut\`onoma de Barcelona,
    Bellaterra E-08193 Barcelona, Spain} } \date{2 April 2013}
\maketitle
\begin{abstract}
We show how excited states in QCD can be profitably used to build up
the Polyakov loop in the fundamental representation at 
temperatures below the hadron--quark-gluon crossover.  The conditions
under which a Hagedorn temperature for the Polyakov loop can be
defined are analyzed.
\end{abstract}

 
\bigskip
 
\section{Introduction}

%The idea of heating up matter to unveil the underlying structure
%displaying quantum effects and the related spectrum is rather old. A
%best and time honoured example was provided by the Debye theory of
%specific heat of solids 100 years ago where acoustic phonons explain
%as collective degrees of freedom much of the observed data without
%need of details on the crystal lattice~\cite{}.  Actually, this
%success makes pinning down improvements rather difficult since
%corrections are most important at low temperatures. 
%Einstein theory of just one simple oscillator required high precision
%low temperature measurements to

The QCD equation of state can be derived from the partition function
\begin{eqnarray}
Z(T)= {\rm Tr} e^{-H/T} = \sum_n g_n e^{-E_n/T} \, . 
\label{eq:ZT}
\end{eqnarray}
In lattice QCD with 2+1 flavours $Z(T)$ has been evaluated by the
HotQCD~\cite{Petreczky:2012rq} and
Wuppertal-Budapest~\cite{Fodor:2012rma} collaborations producing
different results for the trace anomaly at temperatures above $T= 200
{\rm MeV}$, already beyond the hadron--quark-gluon crossover~\cite{Aoki:2006we}.  On the
other hand, Quark-Hadron duality at finite temperature requires that
for confined states $Z$ should be determined from {\it all} stable
hadron states such as those in the PDG
booklet~\cite{Nakamura:2010zzi}. This is the idea behind the Hadron
Resonance Gas (HRG), a multicomponent gas of non-interacting massive
stable and point-like particles~\cite{Hagedorn:1984hz} which has
historically arbitrated the discrepancies between  different
lattice groups~\cite{Karsch:2003zq,Borsanyi:2010cj,Huovinen:2009yb}.
Remarkably, the disagreement still persists beyond the expected range
of validity of the HRG model (see e.g. Fig.~\ref{fig:hw-hag}, right).

The special role played by the HRG does not make it a theorem and
corrections to it are not completely clear as PDG hadrons are
composite, have finite size and width.  On the lattice, the validity
of the HRG has been checked in the strong coupling limit and for heavy
quarks to lowest orders~\cite{Langelage:2010yn}. In the usual large
$N_c$-limit (see Ref.~\cite{Lucini:2012gg} for a review and references
therein) where hadrons become stable resonances, $\Gamma/M= {\cal
  O}(1/N_c)$, the mesons give a finite contribution as their mass and
degeneracy are finite whereas baryons would provide a vanishing
contribution.  The half-width rule~\cite{Arriola:2012vk} applied to
PDG resonances~\cite{Nakamura:2010zzi} provides compatible
uncertainties with current lattice calculations~\cite{Fodor:2012rma}.

To saturate the partition function, Eq.~(\ref{eq:ZT}) with light or
heavy quarks a large number of highly excited states is needed so
relativistic corrections are important. Here, we will use the MIT Bag
model~\cite{Johnson:1975zp} and the Relativized Quark Model (RQM) of
Refs.~\cite{Godfrey:1985xj,Capstick:1986bm} which treats hadrons as
extended bound states rather than resonances.

\section{Trace anomaly and light quarks}

\begin{figure}
\subfigure{\includegraphics[angle=0,width=0.499\textwidth]{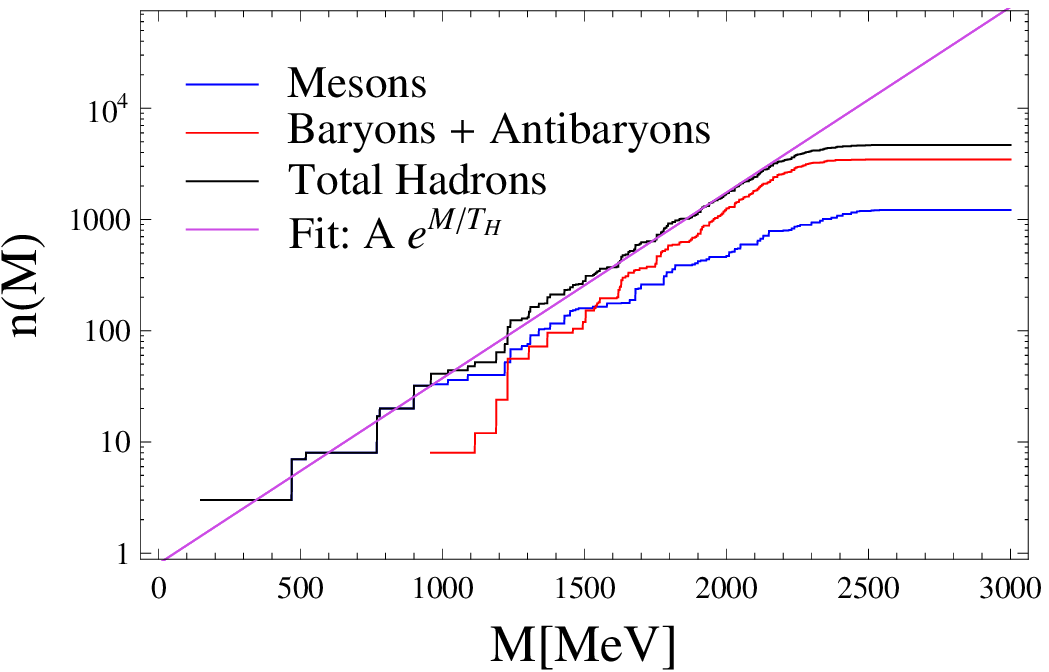}} \hfill
\subfigure{\includegraphics[angle=0,width=0.49\textwidth]{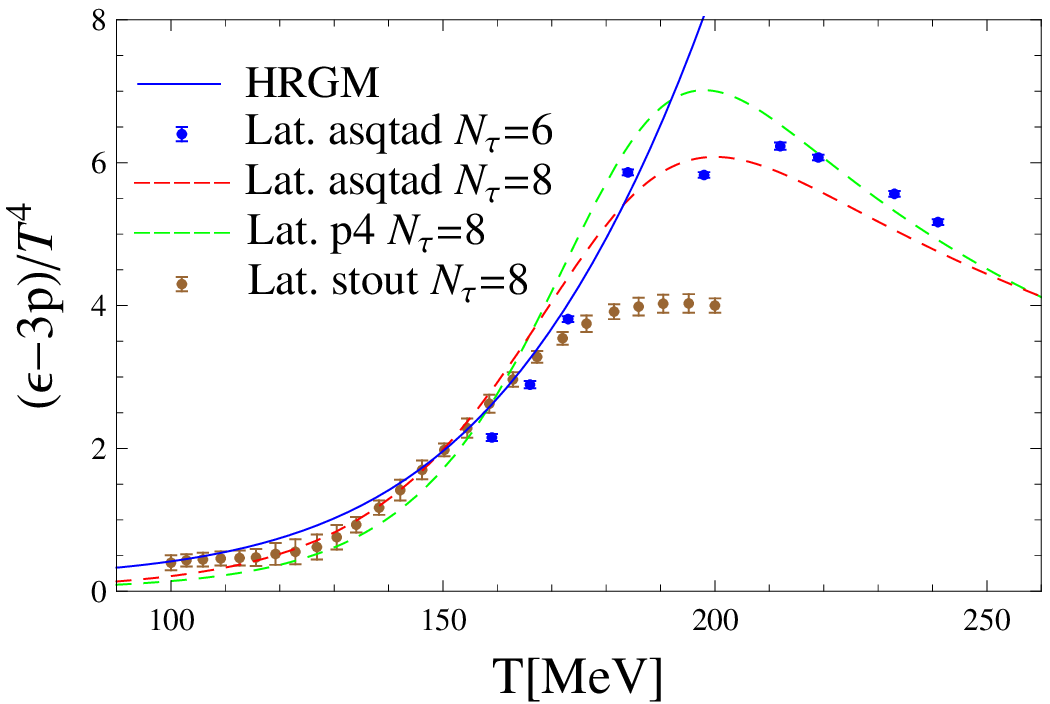}}
\caption{Left: Cumulative number $n$ as a function of the hadron mass
  $M$ (in $\MeV$) with $u$, $d$ and $s$ quarks, computed in the
  RQM~\cite{Godfrey:1985xj,Capstick:1986bm} and compared to a fit
  $n(M)=A e^{M/T_H}$. Right: Trace anomaly $(\epsilon - 3p)/T^4$ as a function of
  temperature (in MeV). We compare lattice data for asqtad and
  p4~\cite{Bazavov:2009zn} (after temperature down-shift of $T_0 =
  15\,\MeV$) and stout~\cite{Borsanyi:2010bp} actions,  with the HRGM
  computed with the RQM spectrum with $u$, $d$ and $s$ quarks from
  Refs.~\cite{Godfrey:1985xj,Capstick:1986bm}.}
\label{fig:hw-hag}
\end{figure}

The trace anomaly measures departures from scale invariance and reads 
\begin{equation}
{\cal A}(T) \equiv \frac{\epsilon - 3p}{T^4} =
T\frac{\partial}{\partial T}\left( \frac{p}{T^4} \right) \, , 
\end{equation}
after using standard thermodynamics relations for the energy density
$\epsilon=E/V$ and pressure $p=- T\log Z/V$. For the HRG
model we have
\begin{eqnarray}
{\cal A}(T) =
\frac1{T^4} \int_0^\infty dM \frac{dn(M)}{dM} 
\int \frac{d^3
  k}{(2\pi)^3} \frac{(E_k-\vec k \cdot \nabla_k E_k )}{e^{E_k/T}\pm 1} \, , 
\end{eqnarray}
where $E_k=\sqrt{k^2+M^2}$ and $\pm$ corresponds to
Fermions/Bosons and 
\begin{equation}
n(M) = \sum_\alpha g_\alpha \Theta( M - M_\alpha ) \, , 
\end{equation}
is the cumulative number ( $\Theta$ is the step function). Hagedorn
proposed that the cumulative number of hadrons in QCD is approximately
and asymptotically given by $n(M) = A \, e^{M/T_H} $ where $A$ is a
constant and $T_H$ is the so called Hagedorn temperature. We show
results in Fig.~\ref{fig:hw-hag} both for $n(M)$ and ${\cal A}(T)$
fitted with $A=0.80$ and $T_H=260 {\rm MeV}$ and also showing the good
performance of the HRG below $T=180{\rm MeV}$ when the RQM is used.

\begin{figure}
\subfigure{\includegraphics[angle=0,width=0.499\textwidth]{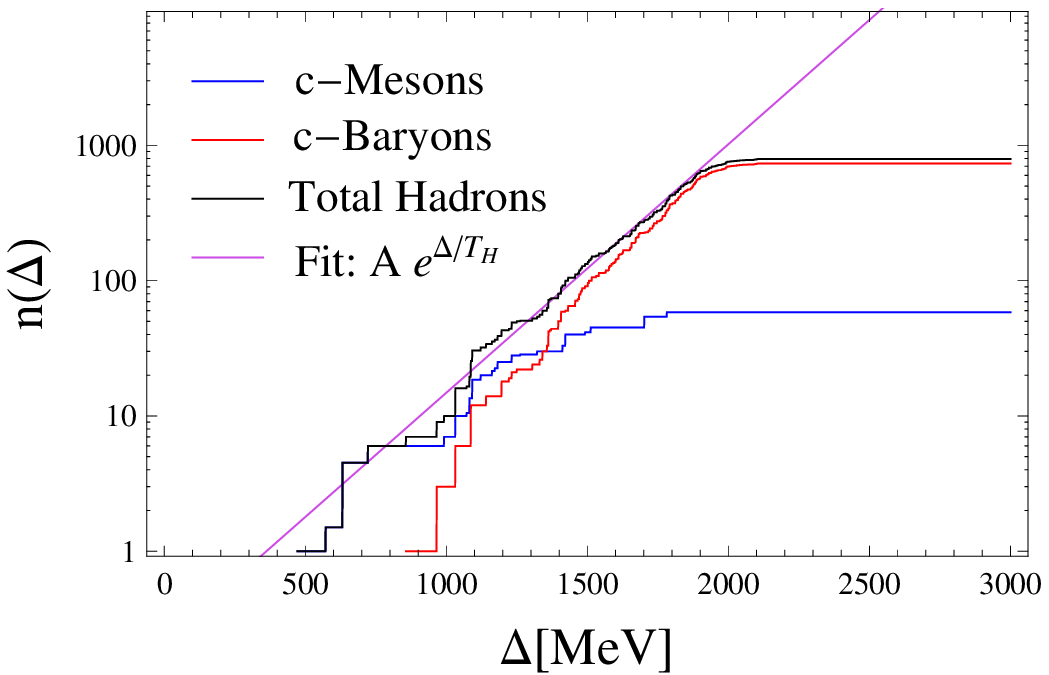}}
\hfill
\subfigure{\includegraphics[angle=0,width=0.49\textwidth]{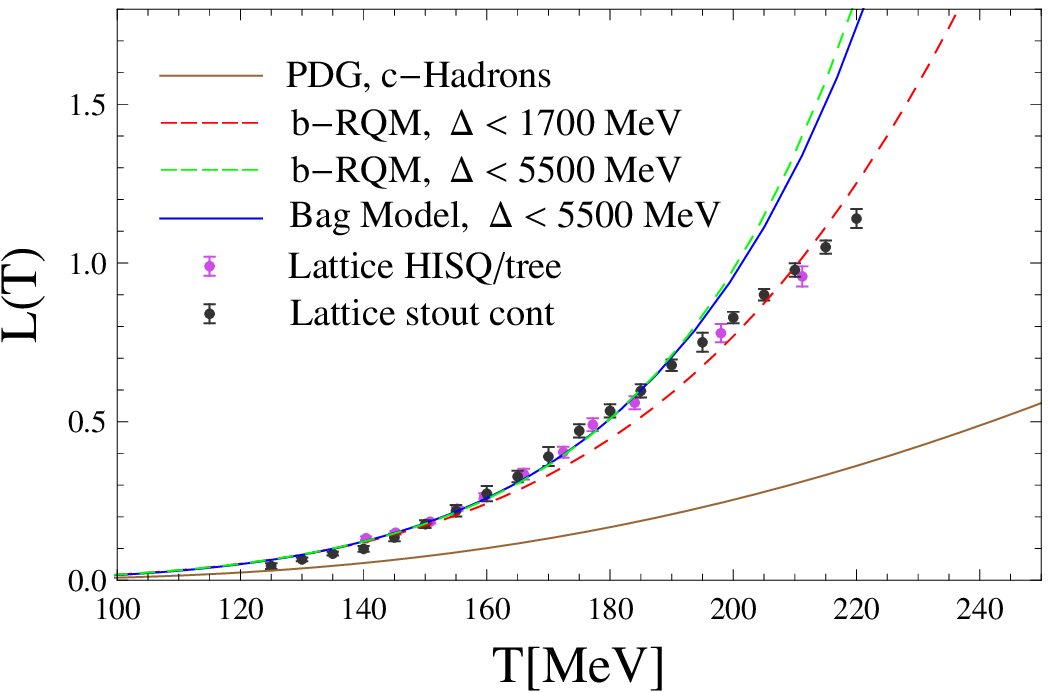}}
\caption{Left: Cumulative number $n$ as a function of the $c$-quark
  mass subtracted hadron mass $\Delta= M-m_c$ (in $\MeV$) with $u$,
  $d$ and $s$ quarks, computed in the
  RQM~\cite{Godfrey:1985xj,Capstick:1986bm} and compared to a fit
  $n(\Delta)=A e^{\Delta/T_H}$. Right: Polyakov loop as a function of
  temperature (in MeV). Lattice data from~\cite{Bazavov:2011nk} for
  the HISQ/tree action and~\cite{Borsanyi:2010bp} for the continuum
  extrapolated stout result. We compare lowest-lying charmed hadrons
  from PDG~\cite{Nakamura:2010zzi}, the RQM spectrum with one $b$
  quark and a cut-off $\Delta < 1700 \,\MeV$ (red line), and $\Delta <
  5500 \, \MeV$ and the MIT bag model ($m_h \to \infty$) with cut-off
  $\Delta < 5500\, \MeV$ is shown as a solid (blue)
  line~\cite{Megias:2012kb}. }
\label{fig:heavy}
\end{figure}

\section{Polyakov loop and heavy quarks}

The Polyakov loop is a purely gluonic operator, which in gluodynamics
becomes a true order parameter as it signals the breaking of the
center symmetry and deconfinement. Unlike the trace anomaly, there is
lattice consensus on this
observable~\cite{Bazavov:2011nk,Borsanyi:2010bp} so its analysis may
be more credible.  We have shown that in
QCD~\cite{Megias:2012kb,Megias:2012hk} and in chiral quark
models~\cite{RuizArriola:2012wd} a hadronic representation exists and
is given by ($A_0$ is the gluon field)
\begin{equation}
L_T = \langle {\rm tr}_c {\sf P} e^{i \int_0^{1/T} \! A_0\, dx_0 } \rangle = \frac{1}{2} \int d\Delta \frac{\partial n(\Delta)}{\partial \Delta} e^{-\Delta/T} \,,
\end{equation}
where the cumulative number reads now   
\begin{eqnarray}
n ( \Delta ) = \sum_\alpha g_{h\alpha} \Theta( \Delta - \Delta_{\alpha,h}) \, , 
\end{eqnarray}
where $g_{h\alpha}$ are the degeneracies and $\Delta_{h\alpha} =
M_{h\alpha} - m_h$ are the masses of hadrons with exactly one heavy
quark (the mass of the heavy quark itself $m_h$ being subtracted). 

The result with $u$, $d$ and $s$ quarks, computed in the
RQM~\cite{Godfrey:1985xj,Capstick:1986bm} when the large but finite
charmed quark mass, $m_h=m_c$ (using $b$-quarks does not change much),
is taken is presented in Fig.~\ref{fig:heavy}. We have checked that
results are not very sensitive to use bottom quarks instead. 
A fit $n(\Delta) = A e^{\Delta/T_H}$ to the total
contribution produces $A = 0.216,0.209 $ and $T_H = 236,207 {\rm
  MeV}$ for single-charmed, bottom hadrons for the range $1 {\rm GeV} \le
\Delta \le 1.8 {\rm GeV}$.
% and $A = 0.216,0.209 $ and $T_H = 236,207 {\rm
%  MeV}$ for a range $0.5
%       {\rm GeV} \le \Delta \le 1.9 {\rm GeV}$
%. whereas  we get $A = 0.142 $ and
%       $T_H = 222 {\rm MeV}$ for single-charmed hadrons and $A = 0.260
% $ and $T_H = 214 {\rm MeV}$ for single-bottom hadrons.  
The results
from PDG and RQM are multiplied by a factor $L(T)\to e^{C/T}
L(T)$, with $C=25\,\MeV$, which comes from an arbitrariness in the
renormalization.  The sum rule has been implemented on the lattice
recently~\cite{Bazavov:2013yv}.

\section{The non-overlapping condition}

\begin{figure}
\subfigure{\includegraphics[angle=0,width=0.49\textwidth]{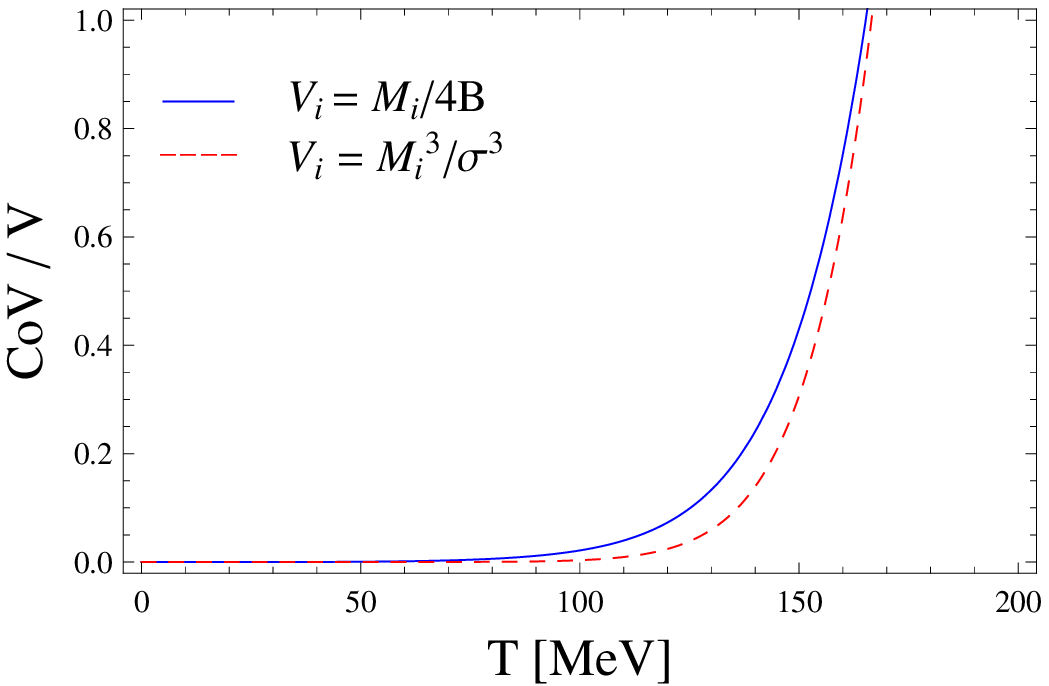}} 
\subfigure{\includegraphics[angle=0,width=0.499\textwidth]{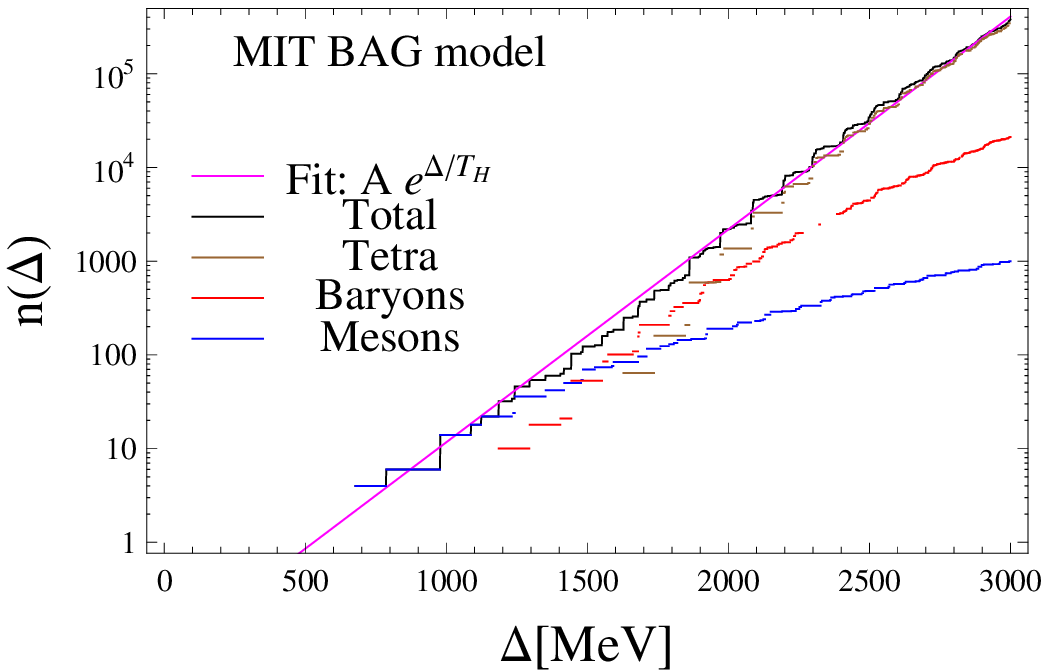}}
\caption{Left: Non-ovelapping condition as a function of temperature.
  For the hadron volume we use $V_i = M_i /4B $ with $B=(0.166 {\rm GeV})^4
  $ for the MIT bag volume (blue) and also $V_i = M_i^3/\sigma^3$ with
  $\sqrt{\sigma}=0.42 {\rm GeV}$ (red). Right: Cumulative number $n
  (\Delta)$ in the MIT Bag model. We include contributions from $Q
  \bar q$, $Q qq$ and $Q \bar q q \bar q $.  }
\label{fig:covol}
%as a function of the heavy-quark subtracted hadron mass
%  $\Delta= \lim (M-m_Q)$ (in $\MeV$)
\end{figure}

In the quantum virial expansion~\cite{Dashen:1969ep}, the excluded
volume corrections come from repulsive interactions whereas resonance
contributions stem from attractive interactions. A good example is
$\pi\pi$ scattering where one has attractive and resonanting states in
the isospin $I=0,1$ corresponding to the $\sigma$ and $\rho$
resonances whereas one has a repulsive core in the $I=2$ exotic
channel~\cite{Venugopalan:1992hy,Kostyuk:2000nx} providing a measure
of the finite pion size. In contrast, the HRG assumes point-like
elementary particles. However, in the narrow width limit resonances
also have a finite size as they become bound states.
%Noting that the particle density reads 
%\begin{eqnarray}
%\frac{N_i}{V} = \int \frac{d^3 p}{(2\pi)^3} \frac{g_i}{e^{E_i(p)/T} \pm 1}
%\end{eqnarray}
%where $E_i(p)=\sqrt{p^2+M_i^2}$. 
Clearly, when hadrons overlap, the HRG model becomes invalid since the
Pauli principle blocks many states allowed by colour neutrality.  The
non-overlapping condition corresponds to the inequality for the Co-volume  
\begin{eqnarray}
{\rm CoV} \equiv\sum _i V_i N_i \le V \, , \qquad \sum_i  V_i \int \frac{d^3 p}{(2\pi)^3} \frac{g_i}{e^{E_i(p)/T} \pm 1} \le 1 \, . 
\end{eqnarray}
The hadron size can be estimated from the MIT bag model where one
has~\cite{Johnson:1975zp} $V_i = M_i/(4B) $.  In the
RQM~\cite{Godfrey:1985xj,Capstick:1986bm} one might compute the size
directly from the m.s.r. of the wave functions.  A meson model of the
form $M=2 p+\sigma r $ with $p \sim 1/r$ yields after minimizing $ V=
4 \pi r^3/3 \sim M^3/\sigma^3 $. In Fig.~\ref{fig:covol} we see
that for $T=160-170 {\rm MeV}$ hadrons overlap and the HRG departs
from the lattice QCD results (see Fig.~\ref{fig:hw-hag}, left).

%In the heavy quark limit we get that $N_i/V \sim e^{-m_Q/T}$ so that
%there are no volume restrictions for the Polyakov loop. 

\section{Hagedorn and the bootstrap}

The cumulative numbers computed in the RQM exhibit lower thresholds
for mesons than baryons but the latter dominate due to the larger
multiplicity of $qqq$ than $q \bar q$ states, and eventually an
exponential growth characterized by a Hagedorn temperature seems to
set in (Figs.~\ref{fig:hw-hag} and \ref{fig:heavy}). Due to the finite
number of degrees of freedom both mesons and baryons have a power-like
behaviour for large masses $M \gg \sqrt{\sigma}$ producing a
dimensional estimate $n_{\bar q q} (M) \sim M^6/\sigma^3$ and $n_{qqq}
(M) \sim M^{12}/\sigma^6$ featuring the available phase space.  An
intriguing issue is under what conditions this exponential growth goes
on high up in the spectrum as initially speculated by
Hagedorn~\cite{Hagedorn:1984hz}. In Fig.~\ref{fig:covol} (right) we
show $n(\Delta)= n_{Q \bar q}(\Delta)+ n_{Q qqq}(\Delta)+ n_{Q \bar q
  q \bar q}(\Delta)+ \dots $ in the MIT Bag model including also the
exotic tetraquark $Q q \bar q \bar q$ states as independent hadronic
states. The fit yields $T_H \sim 191 {\rm MeV}$, complying with the
bootstrap mechanism proposed long
ago~\cite{Kapusta:1981ay,Kapusta:1981ue}. Since some of the tetraquark
states are of molecular nature, it is unclear if they should be
incorporated in the cumulative number. This is related to the
completeness or redundancy of hadronic states, particularly in the PDG
as noted in~\cite{Arriola:2012vk}.

\section{Conclusions}

The thermodynamical analysis of the hadronic spectrum has an
increasing lack of energy resolution for increasing temperatures and a
slowly converging pattern requiring many excited states. On the other
hand, lattice calculations become difficult at very low temperatures
where the main energy gaps are found. While this explains why the HRG
model works well as function of temperature it is not obvious how to
systematically compute deviations from this simple limit.

%\bibliography{refs,refs2}
%\bibliographystyle{h-elsevier}

\end{document}